\documentclass[5p,twocolumn]{elsarticle}
\usepackage{graphicx}
\usepackage{dcolumn}
\usepackage{bm}
\usepackage{hyperref}
\usepackage[mathlines]{lineno}
\usepackage{amsmath}

\usepackage{xcolor}
\usepackage{comment}
\definecolor{Darkgreen}{RGB}{30,120,30}

\begin{document}

\begin{titlepage}
\begin{flushright}
LU TP 19-38\\
August 2019\\
\end{flushright}
\vfill
\begin{center}
{\Large\bf Short-distance constraints for the HLbL contribution to the muon \\[0.25cm]anomalous magnetic moment}\\[2cm]
{\bf Johan Bijnens, Nils Hermansson-Truedsson and Antonio Rodr\'iguez-S\'anchez}
\\[0.5cm]
{Department of Astronomy and Theoretical Physics, Lund University}\\
{S\"olvegatan 14A, SE 223-62 Lund, Sweden}
\vfill
{\bf Abstract}\\[0.5cm]

\begin{minipage}{0.8\textwidth}
We derive short-distance constraints for the hadronic light-by-light
contribution (HLbL) to the anomalous magnetic moment of the muon
in the kinematic
region where the three virtual momenta are all large. We include the external
soft photon via an external field leading to a well-defined
Operator Product Expansion.
We establish that the perturbative quark loop gives the
leading contribution in a well defined expansion.
We compute the first nonzero power correction. It is related to to the
magnetic susceptibility of the QCD vacuum. The results can be used as
model-independent short-distance constraints for the very many different
approaches to the HLbL contribution. Numerically the power correction is found to be small.
\end{minipage}
\end{center}
\vfill
\end{titlepage}
\title{Short-distance constraints for the HLbL contribution to the muon anomalous magnetic moment}

\author[1]{Johan Bijnens}\ead{bijnensthep.lu.se}
\author[1]{Nils Hermansson-Truedsson}%
 \ead{nils.hermansson-truedsson@thep.lu.se}
\author[1]{Antonio Rodr\'iguez-S\'anchez}%
 \ead{antonio.rodriguez@thep.lu.se}
\address[1]{Department of Astronomy and Theoretical Physics, Lund University, S\"olvegatan 14A, SE 223-62 Lund, Sweden.}

\begin{abstract}
We derive short-distance constraints for the hadronic light-by-light
contribution (HLbL) to the anomalous magnetic moment of the muon
in the kinematic
region where the three virtual momenta are all large. We include the external
soft photon via an external field leading to a well-defined
Operator Product Expansion.
We establish that the perturbative quark loop gives the
leading contribution in a well defined expansion.
We compute the first nonzero power correction. It is related to to the
magnetic susceptibility of the QCD vacuum. The results can be used as
model-independent short-distance constraints for the very many different
approaches to the HLbL contribution. Numerically the power correction is found to be small.
\end{abstract}

\maketitle


\section{Introduction}
\label{sec:intro}

The anomalous magnetic moment of the muon is one of the most powerful
low-energy probes of the Standard Model (SM). Its experimental
value via $a_\mu=(g_\mu-2)/2$, \cite{Bennett:2006fi,PhysRevD.98.030001},
\begin{equation}
\label{eq:amuexp}
a^{\textrm{exp}}_{\mu}= 116 \, 592\,  091(63) \times 10^{-11} \, ,
\end{equation}
is expected to be significantly improved \cite{Grange:2015fou,Abe:2019thb}.
The present theoretical prediction is \cite{PhysRevD.98.030001}
\begin{equation}
\label{eq:amutheo}
a_{\mu}^{\textrm{SM}}= 116 \, 591 \, 823(43) \times 10^{-11} \, .
\end{equation}
The tension between (\ref{eq:amuexp}) and
(\ref{eq:amutheo}) might be a sign of physics beyond the SM. Both the theoretical prediction and
the measured value thus need improvement.
Reviews of the theory are \cite{Jegerlehner:2017gek,Jegerlehner:2009ry}.

\begin{figure}[tbh]
\centerline{\includegraphics[width=0.25\textwidth]{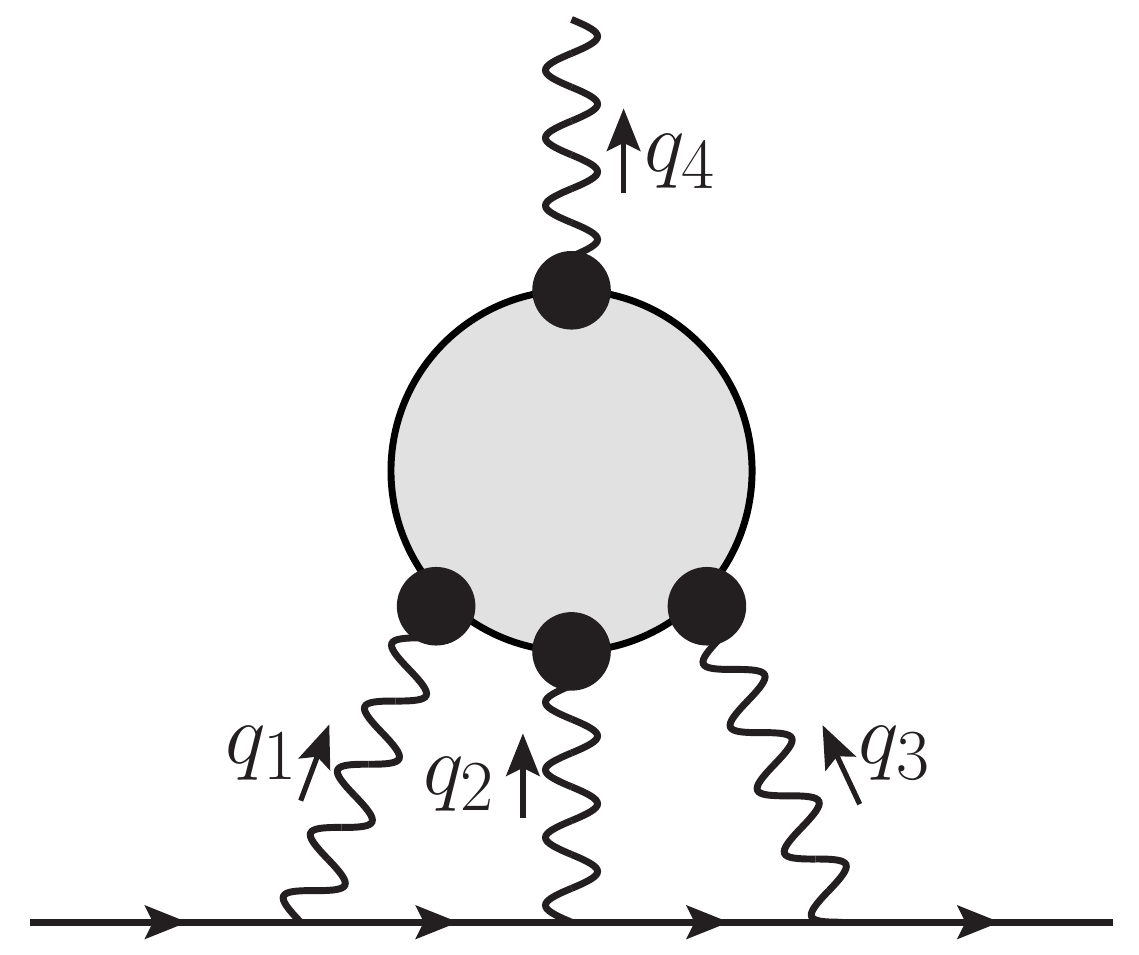}}
\caption{\label{fig:hlbl} The HLbL contribution to the $g-2$. The bottom line is the muon. The blob is filled with hadrons.}
\end{figure}

A major contributor to the theoretical error is the hadronic light-by-light
contribution (HLbL or $a_\mu^\textrm{HLbL}$) depicted in Figure \ref{fig:hlbl}.
It involves the evaluation of the 4-point correlation function of 
electromagnetic quark currents 
\begin{align}\label{eq:hlblten}
    \Pi ^{\mu \nu \lambda \sigma}(q_{1},q_{2},q_{3}) &=\,  -i 
     \int  d^{4}x \, d^{4}y \, d^{4}z \, e^{-i(q_{1}\cdot x +q_{2}\cdot y +q_{3}\cdot z)} 
    \nonumber \\
    & \, \times \langle 0 |T\left\{ J^{\mu}(x) J^{\nu}(y) J^{\lambda}(z) J^{\sigma }(0)\right\} | 0 \rangle \, ,
\end{align}
the HLbL tensor.
The currents are $J^{\mu}(x) = \overline{q}Q_{q}\gamma ^{\mu}q$
with the quark fields $q=(u,d,s)$ and
charge matrix $Q_{q}= \textrm{diag}(2/3,-1/3,-1/3)$. The contribution from the
heavy quarks, $c,b,$ and $t$, can be evaluated fully perturbatively \cite{Kuhn:2003pu}.
The evaluation of $a_\mu^\textrm{HLbL}$ involves an integration with
the loop momenta, $q_1,q_2,$ and $q_3$, running over all possible values and the
fourth, $q_4=q_1+q_2+q_3$, in the static limit,
{\it i.e.} $q_{4}\rightarrow 0$. This class of diagrams
thus contains a complex interplay of strong interactions at different
scales. In the below we  work in the Euclidean domain and use
$Q_i^2=-q_i^2$.

The first full calculations of HLbL were done in the 1990s \cite{Bijnens:1995xf,Hayakawa:1997rq} using mainly models. A model independent approach using dispersive
theory allows for a much more precise determination
\cite{Colangelo:2015ama,Colangelo:2017fiz} for individual
intermediate states but the short-distance part contains
very many. Perturbative short-distance constraints have been used
in constraining individual contributions starting in \cite{Bijnens:1995xf,Knecht:2001qf} as well as some matching with the quark loop~\cite{Bijnens:1995xf}.
The part with $Q_1^2\approx Q_2^2 \gg Q_3^2$ was
treated in \cite{Melnikov:2003xd}.

\begin{figure}[t]
\includegraphics[width=0.23\textwidth]{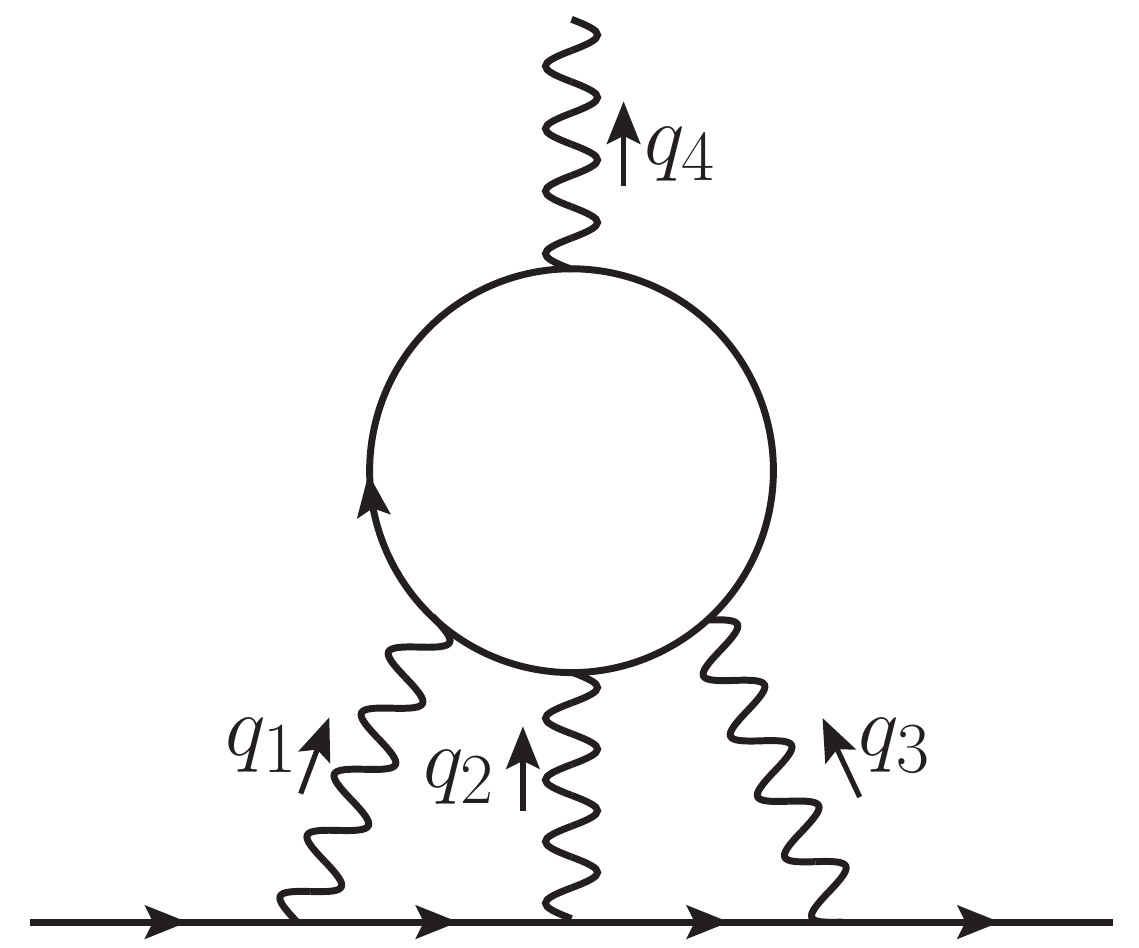}
\includegraphics[width=0.23\textwidth]{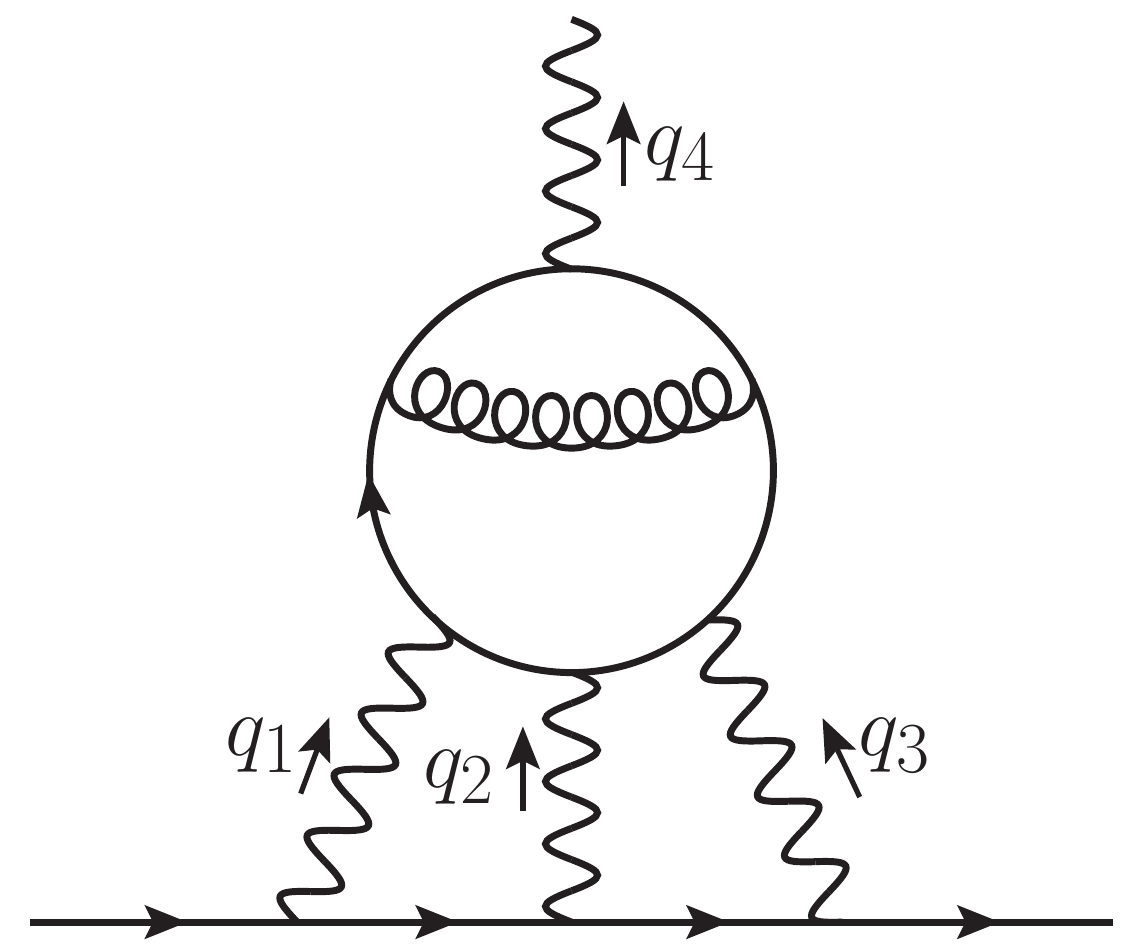}\\[2mm]
\centerline{\includegraphics[width=0.23\textwidth]{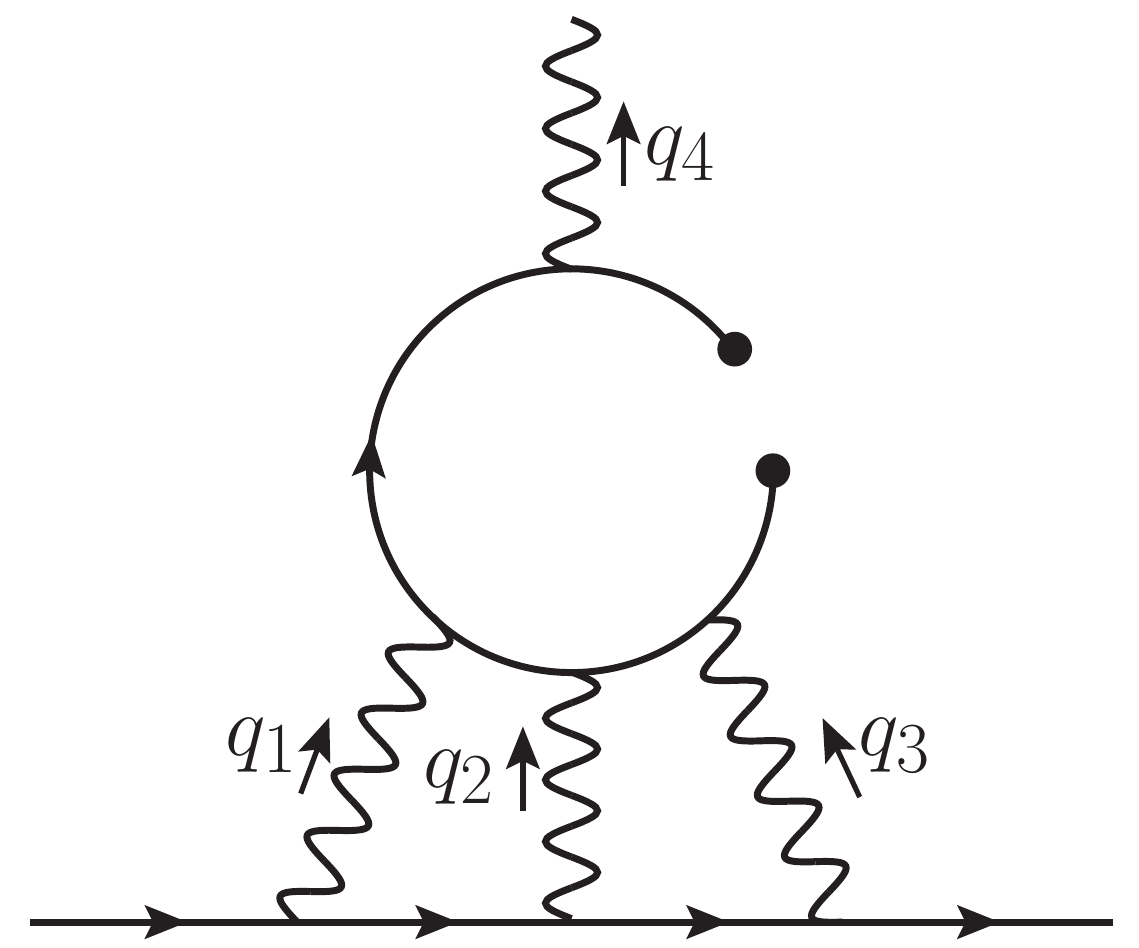}}
\caption{Three examples of short-distance contributions to the HLbL when all $q_{i}$ are large. (a) Pure quark loop, (b) gluonic corrections, (c) contribution from a vev.}\label{fig:hlblsd}
\end{figure}

Our best theoretical understanding of $\Pi^{\mu\nu\lambda\sigma}$ lies in the
kinematic regions where the four Euclidean momenta are large,
$Q_{1}\sim Q_{2} \sim Q_{3} \sim Q_{4}\gg \Lambda_{\textrm{QCD}}$,
where $\Lambda_{\textrm{QCD}}$ is the hadronic scale. This allows for a
perturbative description in terms of quarks and gluons. In this regime, one
may construct a well-defined Operator Product Expansion (OPE), where the
leading contribution corresponds to a simple quark loop with $\alpha_{s}=0$.
Nonperturbative corrections arising from nonzero expectation values of
operators involving quarks and gluons~\cite{Shifman:1978bx}, are suppressed by
powers of $(\Lambda_{\textrm{QCD}}/Q_{i})^{D}$, starting at $D=4$. Some of the
different contributions are sketched in Figure~\ref{fig:hlblsd}. While the
calculation of the different terms of that expansion may be interesting for
constraining some of the models, it does not correspond to any of the
kinematic regions associated with the $g-2$ integral, {\it i.e.} $q_{4}\rightarrow 0$, and we will not discuss this region further.

When considering that the photon associated to the external field should be
set as soft, the OPE mentioned above is no longer valid, even though the three
loop momenta, $Q_{1}$, $Q_{2}$ and $Q_{3}$, are large. This can {\it e.g.} be
seen at the perturbative level when gluonic corrections are considered. 
Setting $\mu\sim Q_{i\neq 4}$, so that $\alpha_{s}(\mu)$ remains small,
one would obtain corrections scaling
as $\alpha^{n}_{s}(Q_{i}) \ln^{m}\frac{Q_{4}}{Q_{i}}$, which break the expansion. The invalidity of the simple OPE for this region becomes even more evident
when trying to compute power corrections such as the one in
Figure~\ref{fig:hlblsd}c. Since no loop momentum flows through the loop, one of the quark propagators depends only on the soft momentum $q_{4}$ and 
is thus manifestly divergent in the static limit.

An analogous problem arises when trying to estimate the nucleon
magnetic moment through the use of baryonic sum rules and it was successfully
solved by formulating an alternative OPE in the presence of an external
electromagnetic background field \cite{Ioffe:1983ju}. 
Note the anomalous magnetic moment is defined classically in an external
magnetic field\footnote{We realized during the course of this work that a similar method has been used for another contribution to $a_\mu^\textrm{HLbL}$
in \cite{Czarnecki:2002nt}.}.
 In this formalism, the soft emission (or, equivalently, response to the
constant external field) can be produced not only by hard quark lines, but
also by low-energy degrees of freedom via vacuum expectation values of
operators. Note that not only operators with vacuum quantum numbers
acquire non-zero values, but also those with the same quantum numbers as
the external electromagnetic field $F_{\mu\nu}$,
{\it e.g.} $\langle \overline{q}\,  \sigma^{\mu\nu}\, q \rangle$.  

In this letter we show how this formalism can be used to provide a model-independent and accurate description of the region where the three incoming loop momenta are large.

\section{Some generalities about the HLbL tensor}

We use the notation of \cite{Colangelo:2015ama,Colangelo:2017fiz} to
facilitate using our results together with theirs. This section summarizes
what we need from there.
The HLbL tensor satisfies the Ward identities
\begin{align}\label{eq:wardid}
    \{ q_{1}^{\mu},q_{2}^{\nu},q_{3}^{\lambda},q_{4}^{\sigma}\} \, \Pi _{\mu \nu \lambda \sigma}(q_{1},q_{2},q_{3}) = 0 \, .
\end{align}
Note that this implies that
\begin{align}\label{eq:wider}
\Pi ^{\mu \nu \lambda\sigma }(q_{1},q_{2},q_{3})=\,&-q_{4 \rho}\frac{\partial\,  \Pi^{\mu\nu\lambda\rho}}{\partial q_{4 \sigma}}(q_{1},q_{2},q_{3}) \, .
\end{align}
The dependence on $q_{4}$ is via $q_{4}=q_{1}+q_{2}+q_{3}$.
Equation~(\ref{eq:wider}) allows to compute $a_\mu^\textrm{HLbL}$ directly from the
derivative \cite{Aldins:1970id}.
In~\cite{Colangelo:2017fiz}, the HLbL tensor is decomposed in a basis with 54 Lorentz scalar functions $\hat\Pi_i$ free of kinematic singularities as
\begin{align}\label{eq:pidecTiHat}
    \Pi ^{\mu \nu \lambda \sigma}(q_{1},q_{2},q_{3}) = \sum _{i=1}^{54}\hat{T}_{i}^{\mu \nu \lambda \sigma} \hat{\Pi} _{i} (q_{1},q_{2},q_{3})\,.
\end{align}
The $\hat{T}_{i}^{\mu \nu \lambda \sigma}$ satisfy Ward identities equivalent to~(\ref{eq:wardid}) and thus, in the static limit $q_4\to0$,
\begin{align}
\label{eq:derivPi}
    & \frac{\partial\,  \Pi^{\mu \nu \lambda \rho}(q_{1},q_{2},q_{3})}{\partial q_{4\sigma }} =
 \sum _{i=1}^{54} \frac{\partial\,  \hat{T}_{i}^{\mu \nu \lambda \rho}(q_{1},q_{2},q_{3})}{\partial q_{4\sigma }} \, \hat{\Pi} _{i}(q_{1},q_{2},q_{3}) \, .
\end{align}
However, in this limit only 19 terms survive~\cite{Colangelo:2017fiz,Melnikov:2003xd}
and using the symmetry $(q_1,\mu)\leftrightarrow(q_2,\nu)$ one obtains
\begin{align}\label{eq:amuint}
    a_{\mu}^{\textrm{HLbL}} = \frac{2\alpha ^{3}}{3\pi ^{2}} 
    & \int _{0}^{\infty} dQ_{1}\int _{0}^{\infty} dQ_{2} \int _{-1}^{1}d\tau \, \sqrt{1-\tau ^{2}}\, Q_{1}^{3}Q_{2}^{3}
    \nonumber \\
    & \times \sum _{i=1}^{12} T_{i}(Q_{1},Q_{2},\tau)\, \overline{\Pi}_{i}(Q_{1},Q_{2},\tau)\, .
\end{align}
The integration variable $\tau$ is defined via $Q_3^2=Q_1^2+Q_2^2+2\tau Q_1Q_2$.
Expressions for the $T_{i}$ can be found in~\cite{Colangelo:2017fiz}, and the $\overline{\Pi}_{i}$ are related to the $\hat{\Pi}_{i}$ according to
\begin{align}
    & \overline{\Pi}_{1} = \hat{\Pi}_{1} \, , \; \overline{\Pi}_{2} = C_{23}\left[  \hat{\Pi}_{1}\right] \, , \; \overline{\Pi}_{3} = \hat{\Pi}_{4} \, , \; \overline{\Pi}_{4} = C_{23}\left[\hat{\Pi}_{4}\right]\, , 
    \nonumber \\
    & \overline{\Pi}_{5} = \hat{\Pi}_{7} \, , \; \overline{\Pi}_{6} = C_{12}\left[ C_{13}\left[  \hat{\Pi}_{7}\right] \right] \, , \; \overline{\Pi}_{7} = C_{23}\left[\hat{\Pi}_{7}\right] \, , \; 
    \nonumber  \\
    & \overline{\Pi}_{8} = C_{13}\left[\hat{\Pi}_{17}\right]\, , \; 
     \overline{\Pi}_{9} = \hat{\Pi}_{17} \, , \; \overline{\Pi}_{10} = \hat{\Pi}_{39} \, , \; 
     \nonumber  \\
    & \overline{\Pi}_{11} = -C_{23}\left[ \hat{\Pi}_{54}\right]  \, , \; \overline{\Pi}_{12} = \hat{\Pi}_{54}\, ,
\end{align}
where $C_{ij}$ permutes the momenta according to $q_{i}\leftrightarrow q_{j}$ for $i,j\in \{1,2,3\}$. As can be seen, only the six functions $\hat{\Pi}_{i}$ for $i \in \{1,4,7,17,39,54\}$ are needed.


\section{The HLBL tensor in an external field}

The HLBL tensor in~(\ref{eq:hlblten}) can be obtained from
\begin{align}
\label{eq:backdyson1}
&\int  d^{4}x \, d^{4}y \, e^{-i(q_{1}\cdot x +q_{2}\cdot y )} 
    \nonumber  \, \langle 0 |T\left\{ J^{\mu}(x) J^{\nu}(y) J^{\lambda}(0) \right\} | \gamma(-q_4) \rangle\nonumber 
\\
&\equiv-\Pi ^{\mu \nu \lambda }(q_{1},q_{2},q_{3})
=i\epsilon_\sigma(-q_4)\Pi ^{\mu \nu \lambda \sigma}(q_{1},q_{2},q_{3})\,,
\end{align}
where we have captured the fourth photon vertex via the matrix element with
a possibly off-shell photon and defined $q_3=q_4-q_1-q_2$.

In the static limit, $q_4\to0$, one can factor out the soft photon part according to
\begin{align}
\label{eq:defPiF}
\nonumber
\Pi ^{\mu \nu \lambda }(q_{1},q_{2},q_{3})&\equiv \Pi ^{\mu \nu \lambda \rho \sigma}_{F}(q_{1},q_{2})\langle 0 | F_{\rho \sigma} \,  |\gamma (-q_4)\rangle
\\&=  i\, q_{4\rho}\epsilon_{\sigma}(-q_4)\Pi ^{\mu \nu \lambda [\rho \sigma]}_{F}(q_{1},q_{2}) \, ,
\end{align}
where $[\rho\sigma]$ indicates antisymmetrization. Combining~(\ref{eq:defPiF}) with~(\ref{eq:wider})
and~(\ref{eq:backdyson1}) one obtains
\begin{equation}\label{eq:equivalencepi}
\lim_{q_{4}\rightarrow 0}\frac{\partial\,  \Pi^{\mu\nu\lambda\rho}}{\partial q_{4\sigma}}(q_{1},q_{2},q_{3}) = \Pi ^{\mu \nu \lambda [\rho \sigma]}_{F}(q_{1},q_{2}) \, .
\end{equation}
The momentum conservation in the static limit reads $q_{1}+q_{2}+q_{3}=0$.
From the above equivalence, (\ref{eq:equivalencepi}), together with~(\ref{eq:derivPi}), it is possible to obtain the required $\hat{\Pi}_{i}$ to
calculate $a_{\mu}^{\textrm{HLbL}}$.

The short-distance quantity, $\Pi ^{\mu \nu \lambda [\rho \sigma]}_{F}(q_{1},q_{2})$
does not depend on the soft-photon momentum and can be calculated directly
using the methods of OPE in an external electromagnetic field of~\cite{Ioffe:1983ju}.
By construction this procedure is free from infrared
divergent propagators. The coupling to an external field can arise in two
different ways, either via a soft insertion on a hard quark line or from
the vacuum expectation values induced by the external electromagnetic field.

In order to simplify calculations we work in the radial gauge for the
external electromagnetic field. This implies to first order, {\it i.e.} in the static
limit,
\begin{align}
    A_{\sigma}(z)=\frac{1}{2}z^{\rho}F_{\rho \sigma}(0) +\ldots \, ,
\end{align}
allowing to calculate immediately in the $q_4=0$ limit.
This gauge is particularly convenient for the soft QCD parts as well, since it
allows to easily expand non-local terms such as
$\langle \overline{q}(x)q(0)\rangle$ into gauge invariant local ones.
This stems from the equivalence between partial derivatives and covariant
derivatives in expansions of fields such as for
instance $q(x) = q(0)+x^{\mu}D_{\mu}q(0)+\ldots$. A pedagogical introduction is
in~\cite{Pascual:1984zb}.

\begin{figure}[t]
\centering
\includegraphics[width=0.4\textwidth]{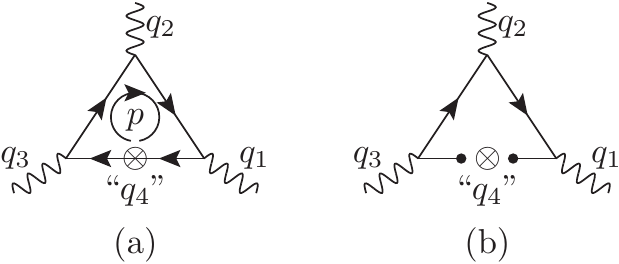}
\caption{\label{fig:opeem} The two leading terms in the external field OPE: (a) The quark loop with loop momentum $p$, and (b) the condensate $\langle \overline{q}\sigma _{\alpha \beta } q\rangle $. The presence of the external field is here represented by a crossed vertex. Note that there is no divergent propagator here as the momentum $q_{4}$ never enters the diagram explicitly.}
\end{figure}

We first look at the contributions with a soft insertion on a hard line.
The lowest order is illustrated in Fig.~\ref{fig:opeem}a.
It is a quark loop with three hard insertions and one soft. The calculation
leads to the same result as the usual quark loop obtained from the calculation
with Fig.~\ref{fig:hlblsd}a, including the dependence on the quark mass.
We have calculated using both methods
as well as compared with quark loop expressions from~\cite{Hoferichter}.
The agreement is exact, both numerical and analytical. In future work we
intend to calculate the gluonic corrections to this. This part shows that the
quark loop at short distances is indeed the first term in a systematic
expansion. We do not quote the analytical expressions since they are rather
lengthy.

We now turn to the power corrections.
The lowest dimensional contribution comes from
\begin{align}\label{eq:magsusc}
    \langle \overline{q}\, \sigma _{\alpha \beta}\, q\rangle \equiv e_{q}F_{\alpha \beta}X_{q} \, ,
\end{align}
where $e_{q}$ is the one of the light quark charges in the matrix $Q_{q}$, and
the $X_{q}$ are so-called tensor coefficients related to the magnetic
susceptilibity that are known from lattice QCD~\cite{Bali:2012jv}.
Regarding the suppression of this condensate as compared to the leading term,
the only scale to compensate dimensions is $\Lambda_{\textrm{QCD}}$. From naive
dimensional analysis, this contribution is thus suppressed by at least a factor
of $\frac{\Lambda_{\textrm{QCD}}}{Q_{\textrm{hard}}}$. 
The contribution is schematically drawn in Fig.~\ref{fig:opeem}b. From
chirality it follows that an extra insertion of a quark mass is needed, so we
get a suppression compared to the quark loop of two powers of the hard scale.
The analytical result for the leading power suppressed contributions are
\begin{align}
    & \hat{\Pi}_{1} = m_{q}X_{q}e_{q}^{4}\, \frac{-4\left( Q_{1}^{2}+Q_{2}^{2}-Q_{3}^{2}\right) }{Q_{1}^{2}Q_{2}^{2}Q_{3}^{4}} \, , & \; \; 
    & \hat{\Pi}_{7} = 0  \, , &
    \nonumber \\
    & \hat{\Pi}_{4} = m_{q}X_{q}e_{q}^{4}\, \frac{8}{Q_{1}^{2}Q_{2}^{2}Q_{3}^{2}} \, , & \; \; 
    & \hat{\Pi}_{39} = 0 \, , &
    \nonumber \\
    & \hat{\Pi}_{17} = m_{q}X_{q}e_{q}^{4}\, \frac{8}{Q_{1}^{2}Q_{2}^{2}Q_{3}^{4}}  \, , & \; \; 
    &  &
      \nonumber \\
    &  \hat{\Pi}_{54} = m_{q}X_{q}e_{q}^{4}\, \frac{-4\left( Q_{1}^{2}-Q_{2}^{2}\right)}{Q_{1}^{4}Q_{2}^{4}Q_{3}^{2}} \, . & &
\end{align}
Work is in progress to calculate the power corrections that are not suppressed
by quark masses but these will be suppressed by more powers of the hard scale.
Preliminary results indicate that the contributions not suppressed by quark
masses occur first suppressed by four powers of the hard scale.

\section{Numerical results}

In this section we present numerical results obtained from the external field OPE. For the numerical integration of $a_{\mu}^{\textrm{HLbL}}$ in~(\ref{eq:amuint}),
 we use the {\sc Cuba} library~\cite{HAHN200578}, both
employing a Monte Carlo algorithm ({\tt Vegas}) as well as a deterministic
algorithm ({\tt Cuhre}) for a cross-check.

First of all we consider the quark loop. In order to compare
with~\cite{Kuhn:2003pu}, we use constituent quark masses
of $m_{u,d,s} = 240$ MeV and $N_{c} = 3$. This yields
$a_{\mu}^{\textrm{HLbL}} = 80.30\times 10^{-11}$, which is in excellent agreement
with the result quoted in~\cite{Kuhn:2003pu}. This was of course expected
given that our leading result analytically agrees with the quark loop.

\begin{figure}[t!]
\centering
\includegraphics[width=0.36\textwidth]{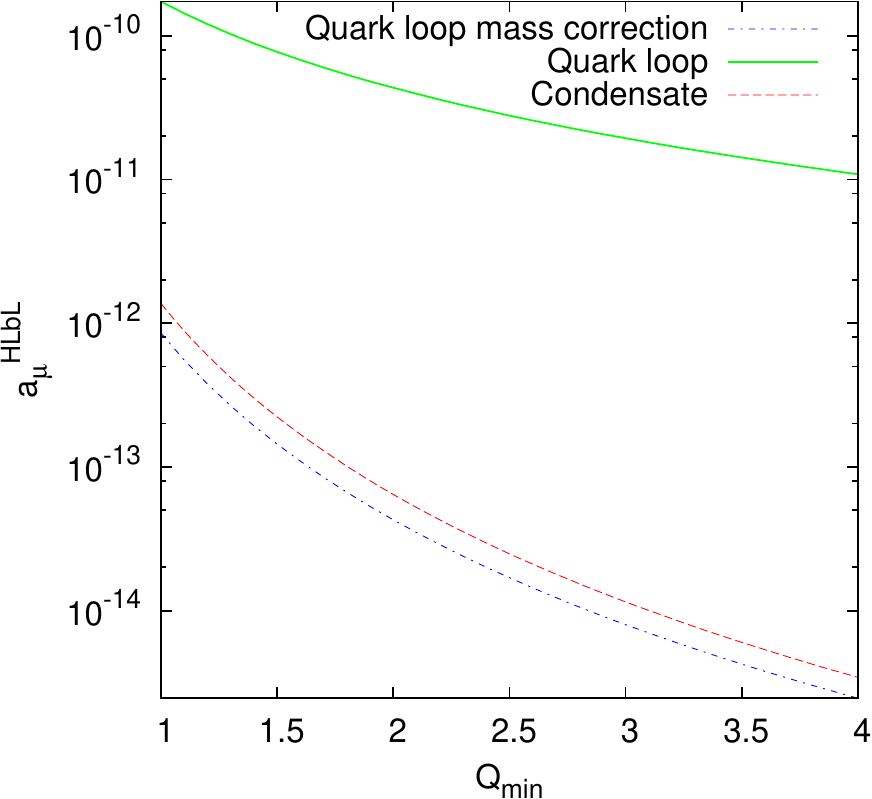}
\caption{\label{fig:running} The $Q_{min}$ dependence of $a_{\mu}^{\textrm{HLbL}}$.}
\end{figure}

We also numerically evaluate the contribution to $a_{\mu}^{\textrm{HLbL}}$
from the regime where our OPE is valid.
In order to allow for future cross-checks, we first calculate it for two
lower cut-offs $Q_{min}= 1,\, 2$ GeV such that $Q_{i=1,2,3}\geq Q_{min}$.
The condensates $X_{q}$ have been estimated in~\cite{Bali:2012jv} on the
lattice, and the values are\footnote{The sign differs from~\cite{Bali:2012jv} due to differences in conventions.}
\begin{align}
    & X_{u} = 40.7 \pm 1.3 \, \textrm{MeV}\, , \; \;  X_{d} = 39.4 \pm 1.4 \, \textrm{MeV}\, , \nonumber \\
    & X_{s} = 53.0 \pm 7.2 \, \textrm{MeV}\, . \; \;
\end{align}
The quark masses we use are $m_{u}=m_{d}= 5$ MeV and $m_{s}=100$ MeV.\footnote{Given the numerical smallness of the result more precise values are not needed.}
The results are presented in Table~\ref{table:condensate}. For an order of
magnitude comparison also the quark loop with zero quark masses is included
there with the same region of integration. As can be seen, the contributions
from the condensates are strongly suppressed
as compared to the quark loop. This is expected given the smallness
of $m_{q}X_{q}$.
\begin{table}[t]
\begin{tabular}{cccc}
$Q_{min}$&Quark Loop&\!$m_{u}X_{u}+m_{d}X_{d}$\!&$m_{s}X_{s}$\\
\hline
\rule{0pt}{4ex}$1$ GeV&$17.3\times 10^{-11}$&$ 5.40\times 10^{-13}$&$ 8.29 \times 10^{-13}$\\
\rule{0pt}{4ex}$2$ GeV&$4.35\times 10^{-11}$&$ 3.40\times 10^{-14}$&$ 5.22\times 10^{-14}$\\
\end{tabular}
\caption{The total contributions to $a^{\textrm{HLbL}}_{\mu}$ from both the quark loop and the next term in the OPE. The condensate contributions have been divided into two parts, one for the up and down quarks and the other for the strange quark.}\label{table:condensate}
\end{table}
\phantom{p}
Finally, in addition to the above comparison we also
look at $a_{\mu}^{\textrm{HLbL}}$ for a range of $Q_{min}$ in
Figure~\ref{fig:running}. The running of the $\overline{\textrm{MS}}$ quark
masses is implemented using the package {\tt CRunDec}~\cite{Schmidt:2012az}. 
In addition to the condensate contribution and massless quark loop, also the
mass correction to the massless quark loop is plotted. As can be seen, both
the condensate contribution and the mass correction scale the same way
in $Q_{min}$. This $Q_{min}$ dependence goes as $1/Q_{min}^{4}$, while the
massless quark loop scales perfectly as $1/Q_{min}^{2}$. 
\phantom{p}
Note that higher-dimensional contributions contain condensates not
suppressed by the small quark mass values. They are expected to dominate the
power corrections when the cut-off $Q_{min}$ is small enough.

\section{Conclusions and outlook}

Due to the long-standing deviation between the experimental value and
the Standard Model prediction of the muon magnetic moment, there is at
present much work going into reducing the errors on both quantities. One of
the two main uncertainties in the Standard Model value comes from the HLbL
contribution, $a_{\mu}^{\textrm{HLbL}}$. The loop integral in
$a_{\mu}^{\textrm{HLbL}}$ is particularly complicated due to the various regions
of virtual or internal photon momenta. In this letter we have focused on
the region where the three (Euclidean) virtual photon momenta are large. 
We have shown how the standard OPE in the vacuum of the associated four-point
correlation function breaks down beyond the leading order in the static limit
in which $g-2$ is defined. Instead, an OPE in the presence of an
electromagnetic background field has been used. The photon associated to the
soft momentum $q_{4}\rightarrow 0$ can be emitted from both high-energy degrees
of freedom, {\it i.e.} quarks, or from long distance ones parametrized by induced
vacuum expectation values of QCD operators. 

The leading order contribution
arises from the radiation of a hard line and is analytically identical to the
purely perturbative quark loop. This proves the expectation that the
perturbative quark loop is the first term in a systematic expansion in this
region. The first power correction in our OPE contains a condensate related
to the magnetic susceptibility of the QCD vacuum. Our numerical study has
shown that its contribution to $a_{\mu}^{\textrm{HLbL}}$ is suppressed, as
compared to the quark loop, by roughly three orders of magnitude, as a
consequence of the small values of the quark masses and the condensate itself. 
The leading contribution scales as suppressed by two powers of the hard scale
while the first power correction is suppressed by four powers of the hard scale.

The higher order power corrections are not suppressed by the small quark
masses. Together with the purely perturbative $\alpha_{s}$ correction, they
should be enough to give a first reliable estimate of the onset of the
asymptotic domain. Both calculations are underway and are expected to be
presented in a forthcoming publication. 


\section*{Acknowledgments}
We thank Martin Hoferichter and Peter Stoffer for useful discussions
and the sharing of their analytic quark loop expressions. 
This work is supported in part by the Swedish Research Council grants contract
    numbers 2015-04089 and 2016-05996, and by the European Research Council
    (ERC) under the European Union's Horizon 2020 research and innovation
    programme under grant agreement No 668679.


\begin{thebibliography}{10}
\expandafter\ifx\csname url\endcsname\relax
  \def\url#1{\texttt{#1}}\fi
\expandafter\ifx\csname urlprefix\endcsname\relax\def\urlprefix{URL }\fi
\expandafter\ifx\csname href\endcsname\relax
  \def\href#1#2{#2} \def\path#1{#1}\fi

\bibitem{Bennett:2006fi}
G.~W. Bennett, et~al., {Final Report of the Muon E821 Anomalous Magnetic Moment
  Measurement at BNL}, Phys. Rev. D73 (2006) 072003.
\newblock \href {http://arxiv.org/abs/hep-ex/0602035}
  {\path{arXiv:hep-ex/0602035}}, \href
  {http://dx.doi.org/10.1103/PhysRevD.73.072003}
  {\path{doi:10.1103/PhysRevD.73.072003}}.

\bibitem{PhysRevD.98.030001}
M.~Tanabashi, et~al.,
  \href{https://link.aps.org/doi/10.1103/PhysRevD.98.030001}{Review of particle
  physics}, Phys. Rev. D 98 (2018) 030001.
\newblock \href {http://dx.doi.org/10.1103/PhysRevD.98.030001}
  {\path{doi:10.1103/PhysRevD.98.030001}}.
\newline\urlprefix\url{https://link.aps.org/doi/10.1103/PhysRevD.98.030001}

\bibitem{Grange:2015fou}
J.~Grange, et~al., {Muon (g-2) Technical Design Report}\href
  {http://arxiv.org/abs/1501.06858} {\path{arXiv:1501.06858}}.

\bibitem{Abe:2019thb}
M.~Abe, et~al., {A New Approach for Measuring the Muon Anomalous Magnetic
  Moment and Electric Dipole Moment}, PTEP 2019~(5) (2019) 053C02.
\newblock \href {http://arxiv.org/abs/1901.03047} {\path{arXiv:1901.03047}},
  \href {http://dx.doi.org/10.1093/ptep/ptz030}
  {\path{doi:10.1093/ptep/ptz030}}.

\bibitem{Jegerlehner:2017gek}
F.~Jegerlehner, {The Anomalous Magnetic Moment of the Muon}, Springer Tracts
  Mod. Phys. 274 (2017) pp.1--693.
\newblock \href {http://dx.doi.org/10.1007/978-3-319-63577-4}
  {\path{doi:10.1007/978-3-319-63577-4}}.

\bibitem{Jegerlehner:2009ry}
F.~Jegerlehner, A.~Nyffeler, {The Muon g-2}, Phys. Rept. 477 (2009) 1--110.
\newblock \href {http://arxiv.org/abs/0902.3360} {\path{arXiv:0902.3360}},
  \href {http://dx.doi.org/10.1016/j.physrep.2009.04.003}
  {\path{doi:10.1016/j.physrep.2009.04.003}}.

\bibitem{Kuhn:2003pu}
J.~H. Kuhn, A.~I. Onishchenko, A.~A. Pivovarov, O.~L. Veretin, {Heavy mass
  expansion, light by light scattering and the anomalous magnetic moment of the
  muon}, Phys. Rev. D68 (2003) 033018.
\newblock \href {http://arxiv.org/abs/hep-ph/0301151}
  {\path{arXiv:hep-ph/0301151}}, \href
  {http://dx.doi.org/10.1103/PhysRevD.68.033018}
  {\path{doi:10.1103/PhysRevD.68.033018}}.

\bibitem{Bijnens:1995xf}
J.~Bijnens, E.~Pallante, J.~Prades, {Analysis of the hadronic light by light
  contributions to the muon g-2}, Nucl. Phys. B474 (1996) 379--420.
\newblock \href {http://arxiv.org/abs/hep-ph/9511388}
  {\path{arXiv:hep-ph/9511388}}, \href
  {http://dx.doi.org/10.1016/0550-3213(96)00288-X}
  {\path{doi:10.1016/0550-3213(96)00288-X}}.

\bibitem{Hayakawa:1997rq}
M.~Hayakawa, T.~Kinoshita, {Pseudoscalar pole terms in the hadronic light by
  light scattering contribution to muon g - 2}, Phys. Rev. D57 (1998) 465--477,
  [Erratum: Phys. Rev.D66,019902(2002)].
\newblock \href {http://arxiv.org/abs/hep-ph/9708227}
  {\path{arXiv:hep-ph/9708227}}, \href
  {http://dx.doi.org/10.1103/PhysRevD.57.465, 10.1103/PhysRevD.66.019902}
  {\path{doi:10.1103/PhysRevD.57.465, 10.1103/PhysRevD.66.019902}}.

\bibitem{Colangelo:2015ama}
G.~Colangelo, M.~Hoferichter, M.~Procura, P.~Stoffer, {Dispersion relation for
  hadronic light-by-light scattering: theoretical foundations}, JHEP 09 (2015)
  074.
\newblock \href {http://arxiv.org/abs/1506.01386} {\path{arXiv:1506.01386}},
  \href {http://dx.doi.org/10.1007/JHEP09(2015)074}
  {\path{doi:10.1007/JHEP09(2015)074}}.

\bibitem{Colangelo:2017fiz}
G.~Colangelo, M.~Hoferichter, M.~Procura, P.~Stoffer, {Dispersion relation for
  hadronic light-by-light scattering: two-pion contributions}, JHEP 04 (2017)
  161.
\newblock \href {http://arxiv.org/abs/1702.07347} {\path{arXiv:1702.07347}},
  \href {http://dx.doi.org/10.1007/JHEP04(2017)161}
  {\path{doi:10.1007/JHEP04(2017)161}}.

\bibitem{Knecht:2001qf}
M.~Knecht, A.~Nyffeler, {Hadronic light by light corrections to the muon g-2:
  The Pion pole contribution}, Phys. Rev. D65 (2002) 073034.
\newblock \href {http://arxiv.org/abs/hep-ph/0111058}
  {\path{arXiv:hep-ph/0111058}}, \href
  {http://dx.doi.org/10.1103/PhysRevD.65.073034}
  {\path{doi:10.1103/PhysRevD.65.073034}}.

\bibitem{Melnikov:2003xd}
K.~Melnikov, A.~Vainshtein, {Hadronic light-by-light scattering contribution to
  the muon anomalous magnetic moment revisited}, Phys. Rev. D70 (2004) 113006.
\newblock \href {http://arxiv.org/abs/hep-ph/0312226}
  {\path{arXiv:hep-ph/0312226}}, \href
  {http://dx.doi.org/10.1103/PhysRevD.70.113006}
  {\path{doi:10.1103/PhysRevD.70.113006}}.

\bibitem{Shifman:1978bx}
M.~A. Shifman, A.~I. Vainshtein, V.~I. Zakharov, {QCD and Resonance Physics.
  Theoretical Foundations}, Nucl. Phys. B147 (1979) 385--447.
\newblock \href {http://dx.doi.org/10.1016/0550-3213(79)90022-1}
  {\path{doi:10.1016/0550-3213(79)90022-1}}.

\bibitem{Ioffe:1983ju}
B.~L. Ioffe, A.~V. Smilga, {Nucleon Magnetic Moments and Magnetic Properties of
  Vacuum in QCD}, Nucl. Phys. B232 (1984) 109--142.
\newblock \href {http://dx.doi.org/10.1016/0550-3213(84)90364-X}
  {\path{doi:10.1016/0550-3213(84)90364-X}}.

\bibitem{Czarnecki:2002nt}
A.~Czarnecki, W.~J. Marciano, A.~Vainshtein, {Refinements in electroweak
  contributions to the muon anomalous magnetic moment}, Phys. Rev. D67 (2003)
  073006, [Erratum: Phys. Rev.D73,119901(2006)].
\newblock \href {http://arxiv.org/abs/hep-ph/0212229}
  {\path{arXiv:hep-ph/0212229}}, \href
  {http://dx.doi.org/10.1103/PhysRevD.67.073006, 10.1103/PhysRevD.73.119901}
  {\path{doi:10.1103/PhysRevD.67.073006, 10.1103/PhysRevD.73.119901}}.

\bibitem{Aldins:1970id}
J.~Aldins, T.~Kinoshita, S.~J. Brodsky, A.~J. Dufner, {Photon - photon
  scattering contribution to the sixth order magnetic moments of the muon and
  electron}, Phys. Rev. D1 (1970) 2378.
\newblock \href {http://dx.doi.org/10.1103/PhysRevD.1.2378}
  {\path{doi:10.1103/PhysRevD.1.2378}}.

\bibitem{Pascual:1984zb}
P.~Pascual, R.~Tarrach, {QCD: renormalization for the practitioner}, Lect.
  Notes Phys. 194 (1984) 1--277.

\bibitem{Hoferichter}
M.~Hoferichter, P.~Stoffer, {private communication}.

\bibitem{Bali:2012jv}
G.~S. Bali, F.~Bruckmann, M.~Constantinou, M.~Costa, G.~Endrodi, S.~D. Katz,
  H.~Panagopoulos, A.~Schafer, {Magnetic susceptibility of QCD at zero and at
  finite temperature from the lattice}, Phys. Rev. D86 (2012) 094512.
\newblock \href {http://arxiv.org/abs/1209.6015} {\path{arXiv:1209.6015}},
  \href {http://dx.doi.org/10.1103/PhysRevD.86.094512}
  {\path{doi:10.1103/PhysRevD.86.094512}}.

\bibitem{HAHN200578}
T.~Hahn,
  \href{http://www.sciencedirect.com/science/article/pii/S0010465505000792}{Cuba--a
  library for multidimensional numerical integration}, Computer Physics
  Communications 168~(2) (2005) 78 -- 95.
\newblock \href {http://dx.doi.org/https://doi.org/10.1016/j.cpc.2005.01.010}
  {\path{doi:https://doi.org/10.1016/j.cpc.2005.01.010}}.
\newline\urlprefix\url{http://www.sciencedirect.com/science/article/pii/S0010465505000792}

\bibitem{Schmidt:2012az}
B.~Schmidt, M.~Steinhauser, {CRunDec: a C++ package for running and decoupling
  of the strong coupling and quark masses}, Comput. Phys. Commun. 183 (2012)
  1845--1848.
\newblock \href {http://arxiv.org/abs/1201.6149} {\path{arXiv:1201.6149}},
  \href {http://dx.doi.org/10.1016/j.cpc.2012.03.023}
  {\path{doi:10.1016/j.cpc.2012.03.023}}.

\end{thebibliography}
\providecommand{\noopsort}[1]{}\providecommand{\singleletter}[1]{#1}%

\end{document}